# MCAD: Multi-modal Conditioned Adversarial Diffusion Model for High-Quality PET Image Reconstruction

Jiaqi Cui[1], Xinyi Zeng[1], Pinxian Zeng[1], Bo Liu[2], Xi Wu[3], Jiliu Zhou[1,2], Yan Wang[1(✉)]

[1] School of Computer Science, Sichuan University, China
`wangyanscu@hotmail.com`
[2] The Hong Kong Polytechnic University, Hung Hom, Hong Kong S.A.R.
[3] School of Computer Science, Chengdu University of Information Technology, China

**Abstract.** Radiation hazards associated with standard-dose positron emission tomography (SPET) images remain a concern, whereas the quality of low-dose PET (LPET) images fails to meet clinical requirements. Therefore, there is great interest in reconstructing SPET images from LPET images. However, prior studies focus solely on image data, neglecting vital complementary information from other modalities, e.g., patients' clinical tabular, resulting in compromised reconstruction with limited diagnostic utility. Moreover, they often overlook the semantic consistency between real SPET and reconstructed images, leading to distorted semantic contexts. To tackle these problems, we propose a novel Multi-modal Conditioned Adversarial Diffusion model (MCAD) to reconstruct SPET images from multi-modal inputs, including LPET images and clinical tabular. Specifically, our MCAD incorporates a Multi-modal conditional Encoder (Mc-Encoder) to extract multi-modal features, followed by a conditional diffusion process to blend noise with multi-modal features and gradually map blended features to the target SPET images. To balance multi-modal inputs, the Mc-Encoder embeds Optimal Multi-modal Transport co-Attention (OMTA) to narrow the heterogeneity gap between image and tabular while capturing their interactions, providing sufficient guidance for reconstruction. In addition, to mitigate semantic distortions, we introduce the Multi-Modal Masked Text Reconstruction ($M^3$TRec), which leverages semantic knowledge extracted from denoised PET images to restore the masked clinical tabular, thereby compelling the network to maintain accurate semantics during reconstruction. To expedite the diffusion process, we further introduce an adversarial diffusive network with a reduced number of diffusion steps. Experiments show that our method achieves the state-of-the-art performance both qualitatively and quantitatively.

## 1 Introduction

Positron emission tomography (PET) is an advanced nuclear imaging technique that enables the visualization and quantification of metabolic activity in the human body [1, 2]. In clinic, standard dose PET (SPET) images are required by physicians for their high image quality and rich diagnostic information. However, the high radiation exposure associated with SPET imaging raises health concerns [3]. One solution is to reduce the injected tracer dose, but this can induce unintended noise and artifacts, leading to degraded image quality with limited diagnostic value. To tackle this dilemma, there is great interest in reconstructing SPET images from LPET images, thereby obtaining clinically acceptable PET images while minimizing potential radiation hazards.

In recent years, deep learning has been widely used in high-quality SPET image reconstruction. Several studies [4-7] leverage convolutional neural networks (CNN)-based frameworks to generate SPET images. More recently, generative adversarial networks (GANs) have gained increasing popularity in PET reconstruction for their talent in image synthesis [8-12]. While powerful, GAN-based models typically employ a rapid one-shot sampling without intermediate steps, which inherently limits their abilities to model the complex noise distribution in PET images [13], causing over-smoothing. As a promising alternative, diffusion probabilistic models (DPMs) [14, 15] have emerged for image reconstruction. Compared to GAN-based models, DPM-based methods adopt a gradual sampling process to enhance noise modeling capabilities, thus facilitating a more reliable LPET-to-SPET mapping [22]. For example, Gong *et al.* [16] first explored the potential of DPMs for PET reconstruction. Following this, Han *et al.* [17] designed a DPM-based framework to generate SPET images in a coarse-to-fine manner. Meanwhile, Jiang *et al.* [18] and Shen *et al.* [19] respectively employed a latent-based conditional DPM and a score-based DPM for unsupervised reconstruction.

Despite their satisfactory performance, current methods still face two limitations. First, they concentrate exclusively on leveraging image data while ignoring valuable information present in the patient's non-imaging clinical tabular. The high-level attributes in the tabular data (e.g., age, weight, and sex) can provide crucial complementary guidance for PET reconstruction, given their close correlation with metabolic activities of the human body that PET imaging aims to reveal [20, 21]. Therefore, neglecting such vital information unavoidably results in compromised reconstruction with limited diagnostic clues. Second, they often overlook the high-level semantic consistency between the real SPET and the reconstructed images. Particularly, for DPM-based approaches, the random noise in the diffusion process inherently undermines semantic preservation [22], resulting in reconstructed images with inaccurate semantic contexts.

To address the above limitations and also inspired by the noise modeling capability of DPMs, in this paper, we propose a novel conditional diffusion model, namely MCAD, to reconstruct high-quality SPET images from LPET images and clinical tabular data. Different from prior methods that rely solely on image data, we incorporate multi-modal conditions, utilizing both LPET images to fundamentally direct the reconstruction and non-image clinical tabular to offer complementary perspectives for further image quality enhancement. Specifically, our MCAD incorporates a Multi-modal conditional Encoder (Mc-Encoder) to extract multi-modal input features, and then employs a conditional diffusion process to blend noise with multi-modal inputs, gradually mapping the blended features to the target SPET images. To fully exploit the information in multi-modal inputs, we further integrate Optimal Multi-modal Transport co-Attention (OMTA) into the Mc-Encoder, aiming to minimize the discrepancy between the image and tabular while modeling their correlations. Moreover, to ensure consistency in high-level semantics that are prone to distortion during reconstruction, we develop the Multi-Modal Masked Text Reconstruction (M³TRec). In M³TRec, high-level attributes (age, weight, sex, *etc.*) from the clinical tabular are initially masked and then restored using semantic knowledge extracted from denoised PET images. As the precise semantics in denoised images can effectively facilitate the accurate recovery of masked clinical attributes, the M³TRec, in turn, enforces the preservation of semantics in reconstruction. In addition, inspired by [23], we adopt an adversarial diffusive network to expedite the diffusion process with a reduced number of diffusion steps.

The contributions of this paper can be summarized as follows. (1) We propose a novel multi-modal conditioned adversarial diffusion model that leverages both LPET images and clinical tabular data for reconstructing high-quality SPET images. *To our knowledge, this work marks a pioneering effort in applying clinical tabular to assist SPET reconstruction from LPET.* (2) To fully exploit multi-modal inputs, we devise an OMTA-embedded Mc-Encoder that aligns the disparity between image and tabular data while capturing their interactions, thereby providing complementary guidance to enhance PET reconstruction. (3) We design the M³TRec to encourage consistency in high-level semantic contexts, preventing semantic distortions in the reconstructed images. (4) Experimental results demonstrate the feasibility and superiority of our method.

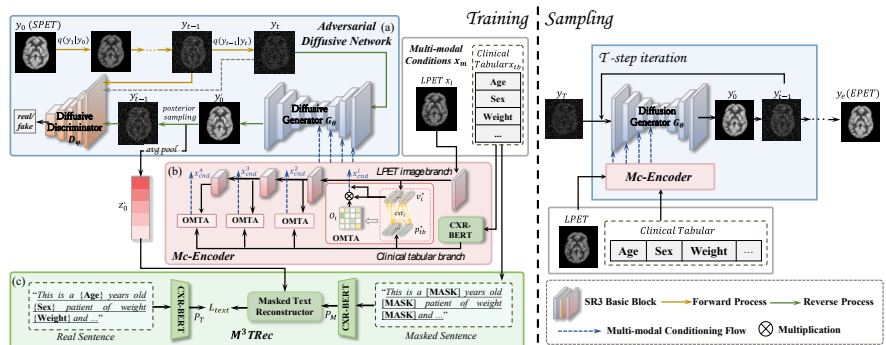

**Fig. 1.** Overview of the proposed MCAD model. During training, (a) the adversarial diffusive network takes the information from multi-modal conditional inputs (i.e., LPET and clinical tabular) that are extracted by (b) Mc-Encoder to generate EPET images. Meanwhile, (c) M³TRec takes clinical tabular to promote consistency in high-level semantics. During sampling, given multi-modal inputs, the well-trained diffusive generator and the Mc-Encoder are utilized to iteratively transform the noise towards EPET images.

## 2 Methodology

The overview of the MCAD is illustrated in Fig. 1, which consists of three parts, i.e., an adversarial diffusive network, a Multi-modal conditional Encoder (Mc-Encoder), and Multi-Modal Masked Text Reconstruction (M³TRec). During training, the adversarial diffusive network unites a diffusive generator and a diffusive discriminator to diffuse on real SPET images, aiming to accelerate the diffusion process. To achieve a more controlled diffusion process and enhance the reconstruction quality, the Mc-Encoder extracts information from multi-modal conditional inputs and hierarchically injects them into the diffusive generator. Meanwhile, the clinical tabular is also fed into the M³TRec to promote consistency in high-level semantics. During sampling, given the LPET image and the clinical tabular, the diffusive generator and the Mc-Encoder are utilized to iteratively transform the noise towards estimated PET (EPET) images.

### 2.1 Adversarial Diffusive Network

As shown in Fig. 1(a), the adversarial diffusive network follows the conditional diffusion framework, involving a forward and a reverse process. In MCAD, we unite a 4-level diffusive generator and a 5-level diffusive discriminator to expedite the reverse sampling process in an adversarial manner.

**Forward Process:** In the forward diffusion process, fixed Gaussian noise is gradually added to SPET $y_0 \sim q(y_0)$ via the Markov chain in $T$ ($T < 10$) steps, producing a sequence of noisy images $\{y_1, y_2, \ldots, y_T\}$, where $y_T \sim \mathcal{N}(0, \mathbf{I})$ is the pure noise. This forward process can be formulated by a predefined variance schedule $\{\beta_1, \beta_2, \ldots, \beta_T\}$ as:

$$q(y_{1:T}|y_0) = \prod_{t=1}^{T} q(y_t|y_{t-1}), \quad q(y_t|y_{t-1}) := \mathcal{N}(y_t; \sqrt{1-\beta_t}y_{t-1}, \beta_t \mathbf{I}), \quad (1)$$

**Reverse Process:** The reverse process aims to recover $y_0$ from $y_T$ by gradually deducing the noise over $T$ time steps. At each time step $t$, provided with the current noisy data $y_t$ and the conditioning information $c$, a neural network parametrized $\theta$ predicts the conditional probability distribution of the denoised version $y_{t-1}$. The predicted distribution $p_\theta(y_{t-1}|y_t, c)$ is expected to approximate its real counterpart $q(y_{t-1}|y_t, c)$. In conventional diffusion models, $p_\theta(y_{t-1}|y_t, c)$ follows a Gaussian distribution, as the step size from $t$ to $t-1$ is sufficiently small to produce a Gaussian distributed $q(y_{t-1}|y_t, c)$. However, when the step size becomes larger, $q(y_{t-1}|y_t, c)$ becomes non-Gaussian [23]. Therefore, for fast reverse sampling (i.e., larger step size), we break the conventional Gaussian assumption and adopt a GAN network that leverages a conditional diffusive generator $G_\theta$ to predict $p_\theta(y_{t-1}|y_t, c)$, and employs a diffusive discriminator $D_\varphi$ to minimize the divergence between the predicted $p_\theta(y_{t-1}|y_t, c)$ and the real distribution $q(y_{t-1}|y_t, c)$. Specifically, $G_\theta$ receives the data pair $(y_t, c)$ as input to predict the non-corrupted image $y'_0$. Then, we utilize the posterior distribution to reintroduce noise to $y'_0$ and sample $y'_{t-1}$ from $p_\theta(y_{t-1}|y_t, c)$ via the following formula:

$$y'_{t-1} \sim p_\theta(y_{t-1}|y_t) := q(y_{t-1}|y_t, y'_0 = G_\theta(y_t, c, t)), \quad (2)$$

For adversarial training, $D_\varphi(\{y'_{t-1} \text{ or } y_{t-1}\}, y_t, t)$ aims to distinguish between samples drawn from the estimated posterior distribution $p_\theta(y_{t-1}|y_t, c)$ (as fake one) and the real counterpart $q(y_{t-1}|y_t, c)$ (as real one). The above process achieves mutual wins in reducing diffusion time steps $T$ and directing the network to capture complex distributions in real SPET images.

### 2.2 Multi-modal Conditional Encoder (Mc-Encoder)

Considering the complex metabolic distributions in PET images, it is challenging to accurately recover SPET images conditioned only on LPET images. Therefore, we introduce multi-modal conditional inputs $x_m = \{x_l, x_{tb}\}$ that contains not only LPET images $x_l$ but also clinical tabular $x_{tb}$ to guide a more precise diffusion process. Particularly, LPET images offers fundamental guidance to direct the reconstruction, whereas the clinical tabular offers complementary metabolic-related perspectives for further image quality refinement. As shown in Fig. 1(b), we design a Mc-Encoder to extract multi-modal features, and then introduce them into the diffusive generator for guidance. To balance the heterogenous modalities of image and tabular, Optimal Multi-modal Transport co-Attention (OMTA) modules are utilized to effectively integrate tabular and image features while inferring their correlations.

**Tabular Branch.** In the tabular branch, clinical tabular $x_{tb} \in \mathbb{R}^L$ are fed to CXR-BERT [25] to generate tabular feature $p_{tb} \in \mathbb{R}^{d \times L}$, where $d$ is the dimension and $L$ is the tabular length. Pre-trained on biomedical texts, CXR-BERT can impart biomedical contexts into $p_{tb}$, enriching them with clinical clues for high-quality reconstruction.

**Image Branch.** Corresponding to the encoding part of $G_\theta$, the image branch also involves 4 levels. Given the LPET image $x_l \in \mathbb{R}^{C \times H \times W}$, multi-level features are extracted and are denoted as $V = \{v_i \in \mathbb{R}^{C_i \times H_i \times W_i}\}_{i=1}^{4}$, where $H_i$, $W_i$, and $C_i$ represent the height, width, and channel dimension of the image feature in $i$-th level, respectively.

**OMTA Module.** After every level in the image branch, an OMTA module is integrated to enable active interactions between image and tabular features. It leverages the matching capability of optimal transport to harmonize the image-tabular discrepancy, while utilizing the selective focusing ability of attention mechanism to emphasize image regions correlated with more relative attributes in the tabular. Specifically, after the $i$-th level in the image branch, image feature $v_i \in \mathbb{R}^{C_i \times H_i \times W_i}$ is flattened to $v_i^* \in \mathbb{R}^{C_i \times M_i}$, where $M_i = H_i \times W_i$. Meanwhile, tabular feature $p_{tb}$ is projected to match the dimension of the flattened image, producing $p_{tb}^* \in \mathbb{R}^{C_i \times L}$. Then, optimal transport between them can be defined by the discrete Kantorovich formulation [26] that searches the overall optimal matching flow $O_i$ between $v_i^*$ and $p_{tb}^*$, which is formulated as:

$$\mathcal{W}(v_i^*, p_{tb}^*) = \min_{M_i \in \prod(\mu_p, \mu_g)} <O_i, cst_i>_F,$$
$$\prod(\mu_p, \mu_g) = \{M_i \in \mathbb{R}_+^{C_i \times L} | O_i \mathbf{1}_L = \mu_p, O_i^T \mathbf{1}_{C_i} = \mu_g\}, \quad (3)$$

where $cst_i \geq 0 \in \mathbb{R}^{C_i \times L}$ is the cost matrix that measures the $l_2$-distance of the image feature $v_i^*$ and the tabular feature $p_{tb}^*$. $\prod(\mu_p, \mu_g)$ describes the distribution between $v_i^*$ and $p_{tb}^*$ based on $cst_i$, under the marginals $\mu_p$ and $\mu_g$. $<\cdot>_F$ is the Frobenius dot product. Once acquiring the optimal matching flow $O_i$, attention mechanism is imposed between $O_i$ and $v_i^*$ to create $\hat{v}_i^* = O_i^T v_i^*$ which contains rich multi-modal information to guide the reconstruction. Finally, $\hat{v}_i^*$ is projected and fused with the image feature $v_i^*$ by concatenation, forming the integrated conditional feature $x_{cnd}^i$. After being reshaped to $\mathbb{R}^{C_i \times H_i \times W_i}$, the $i$-th level conditional feature is added to the corresponding output of the $i$-th level of $G_\theta$. Finally, we can obtain the conditional features of all four levels, denoted as $\{x_{cnd}^i\}_{i=1}^{4}$, which constitute the condition $c$ in Sec. 2.1.

### 2.3 Multi-Modal Masked Text Reconstruction (M³TRec)

To retain essential semantic information during reconstruction, we devise a M³TRec network. Given the clinical tabular $x_{tb}$, we construct a text prompt $P_T$ which is encoded based on the sentence of "This is a {age} years old {sex} patient of weight {weight}…" by CXR-BERT. The curly brackets in the sentence represents the corresponding clinical tabular attributes that contain high-level semantics. Then, we mask these attributes and encode the masked sentence into the masked text prompt $P_M$. Meanwhile, we apply global average pooling to the denoised image $y'_0$ to obtain the semantic embedding $z'_0$. Subsequently, we concatenate $P_M$ with $z'_0$ for joint decoding by the masked text reconstructor $R_M$ to reconstruct masked attributes. Our insight is that accurate semantics captured in $y'_0$ can effectively guide the retrieval of masked attributes through its corresponding global semantic embedding $z'_0$. Therefore, by minimizing the disparity between the reconstructed and the real text prompts during training, we can, in turn, promote the semantic consistency in reconstructed PET images, thus reducing distortions. A masked reconstruction loss is then introduced to supervise the above process.

## 2.4 Sampling

For inference, only the diffusive generator $G_\theta$ and Mc-Encoder are utilized to sample EPET images. Starting from the pure noise $y_T$, we employ $G_\theta$ conditioned on the input LPET images and the clinical tabular to generate $y'_0$. Then, we adopt the posterior sampling in Eq. (2) to get the next step image $y'_{T-1}$. This process last for $T$ iterations. In the $T$-th iteration, $y'_0$ becomes the final EPET image, denoted as $y_e$.

## 2.5 Objective Function

The objective function is comprised of three parts: 1) the image estimation loss, and 2) the masked reconstruction loss, and 3) the adversarial diffusive loss.

The image estimation loss implements L1 loss to minimize the gap between the real SPET image $y_0$ and the EPET image $y_e$, while encouraging less blurring, as follows:
$$L_{image} = E_{y_0, y_e}[||y_0 - y_e||_1], \quad (4)$$

To ensure the semantic consistency, the masked reconstruction loss is computed as the cosine embedding loss $\mathcal{H}(\cdot)$ between embeddings of the real text prompt $P_T$ and its reconstructed counterpart $R_M(P_M, z'_0)$, which is defined as:
$$L_{text} = \mathcal{H}(P_T, R_M(P_M, z'_0)), \quad (5)$$

Following [23], the adversarial diffusive loss can be expressed as:
$$L_{adv} = \min_\varphi \max_\theta \sum_{t \geq 1} E_{q(y_t)} E_{q(y_{t-1}|y_t, c)}[-log(D_\varphi(y_{t-1}, y_t, t)) + E_{p_\theta(y'_{t-1}|y_t, c)}[log(D_\varphi(y'_{t-1}, y_t, t))]], \quad (6)$$

Overall, the final objective function is the weighted sum of the above three losses, which is formulated as below:
$$L_{total} = L_{adv} + \lambda_1 L_{image} + \lambda_2 L_{text}. \quad (7)$$
where $\lambda_1$ and $\lambda_2$ are the weighting coefficients to balance these three terms.

## 3 Experiments

### 3.1 Experimental Settings

**Table 1.** Quantitative comparison results of our MCAD with state-of-the-art PET reconstruction approaches on the UDPET dataset.

| Methods | PSNR [dB]↑ | SSIM↑ | NMSE↓ | Params |
|---|---|---|---|---|
| Auto-Context [3] | 22.762±1.003 | 0.871±0.035 | 0.044±0.014 | 41M |
| TriDo-Former [7] | 22.980±0.389 | 0.878±0.021 | 0.042±0.005 | 38M |
| Ea-GAN [28] | 23.933±0.907 | 0.894±0.015 | 0.034±0.008 | 42M |
| AR-GAN [9] | 24.149±1.023 | 0.894±0.028 | 0.031±0.009 | **8M** |
| CVT-GAN [11] | 24.330±0.969 | 0.901±0.028 | 0.032±0.030 | 16M |
| CDM [16] | 24.393±1.047 | 0.908±0.025 | 0.029±0.008 | 34M |
| **MCAD (Ours)** | **25.080±1.003** | **0.927±0.023** | **0.026±0.007** | 30M |

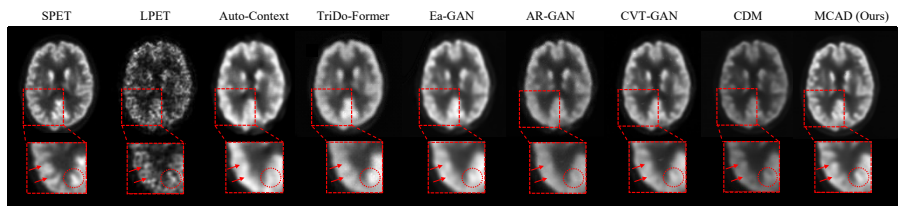

**Fig. 2.** Visual comparison results of images reconstructed by different methods.

**Dataset.** We train and evaluate our proposed method on a public UDPET dataset [27]. In this dataset, 160 subjects of $^{18}$F-FDG PET imaging scanned with Siemens Biograph Vision Quadra are selected, in which 130, 10, and 20 subjects are utilized for training validating, and testing. LPET images are produced by subsampling the SPET scans by a dose reduction factor of 100 to simulate the acquisition at 1/100 of the standard dose. Each 3D scan of brain has a size of 128 × 128 × 128 and is sliced into 2D slices of size 128 × 128. Three standard metrics including peak signal-to-noise (PSNR), structural similarity index (SSIM), and normalized mean squared error (NMSE) are adopted for performance evaluation. We restack 2D slices into whole 3D PET scans for evaluation.
**Implementation Details.** Our network is implemented by Pytorch framework and trained end-to-end on a single GeForce GTX 3090 GPU for 50 epochs with a batch size of 32, using Adam optimizer. For the diffusive adversarial network, each level in $G_\theta$ and $D_\varphi$ consists of two SR3 blocks [24] with modified channel. Similarly, in the Mc-Encoder, each level in the image branch also consists of two SR3 blocks. The initial learning rates for $G_\theta$ and $D_\varphi$ are set to 2e-4 and 1.5e-4, and are adjusted by cosine decay scheduler during training. The diffusion step $T$ is set to 5. The weighting coefficients $\lambda_1$ and $\lambda_2$ in Eq. (7) are empirically set as 100 and 10, respectively.

### 3.2 Comparison with State-of-the-Art Methods

We compare our MCAD against six leading reconstruction methods, including regression-based methods: (1) Auto-Context [4], (2) TriDo-Former [28]; GAN-based methods: (3) Ea-GAN [29], (4) AR-GAN [9], (5) CVT-GAN [11]; and diffusion-based method: (6) CDM [17], on the UDPET dataset. The comparison results are presented in Table 1. As can be seen, our MCAD outperforms the compared methods in all three evaluation criteria with moderate parameters. Particularly, compared with the current leading approach CDM, our proposed method still boosts the PSNR and SSIM by 0.687dB and 0.019, respectively. In addition, our method contains only 30M parameters whereas the CDM has 34M, which further validates the speed and feasibility of our method. Moreover, we conduct the paired t-test and the results show that p-values on three metrics are less than 0.05, suggesting that our improvements are statistically significant. The visual comparison results of our method and the compared approaches are displayed in Fig. 2, where regions with obvious improvements are highlighted by red arrows and circles in the zoomed-in areas. It can be observed that our MCAD yields the best visual effect with the finest details and minimal reconstruction error. Overall, these results suggest the superiority of our MCAD over state-of-the-art methods.

### 3.3 Ablation studies

To verify the effectiveness of the key components of our MCAD, we conduct the ablation study with the following variants: (1) conventional DPM ($T = 1000$) conditioned only on LPET images (i.e., Baseline), (2) introducing adversarial training (i.e., Baseline + Adv), (3) concatenating clinical tabular to LPET images (i.e., Baseline + Adv + Text), (4) utilizing conventional attention (CA) to integrate LPET images and the clinical tabular (i.e., Baseline + Adv + Text + CA), (5) replacing CA with our OMTA (i.e., Baseline + Adv + Text + OMTA), (6) injecting M³TRec (i.e., Proposed). The qualitative results are displayed in Fig. 3. Compared to other ablation variants, it is evident that images reconstructed by the proposed method are the closest to real SPET images and retain richer details, particularly in the areas highlighted by the red boxes. Quantitative results are given in Table 2. It can be observed that the performance of our model progressively improves with the introduction of each component. Particularly, the introduction of the clinical tabular yields a notable 0.605dB improvement in PSNR, indicating its potency in providing vital complementary information to enhance the reconstruction quality. In addition, our OMTA gains superior performance compared to the conventional CA, demonstrating its capability of aligning and modeling the interaction between heterogenous image and tabular data. Moreover, the incorporation of the M$^3$TRec improves the model performance by a large margin, affirming its efficacy in preserving semantic information.

**Table 2.** Quantitative results of the ablation study on the UDPET dataset.

| Adv. | Text | CA | OMTA | Rec | PSNR [dB]↑ | SSIM↑ | NMSE↓ |
|---|---|---|---|---|---|---|---|
| | | | | | 23.477±1.498 | 0.878±0.033 | 0.043±0.018 |
| ✓ | | | | | 23.495±1.350 | 0.878±0.036 | 0.041±0.021 |
| ✓ | ✓ | | | | 24.100±1.028 | 0.883±0.041 | 0.033±0.014 |
| ✓ | ✓ | ✓ | | | 24.364±1.321 | 0.890±0.029 | 0.032±0.012 |
| ✓ | ✓ | | ✓ | | 24.573±1.036 | 0.911±0.025 | 0.029±0.009 |
| ✓ | ✓ | | ✓ | ✓ | **25.080±1.003** | **0.927±0.023** | **0.026±0.007** |

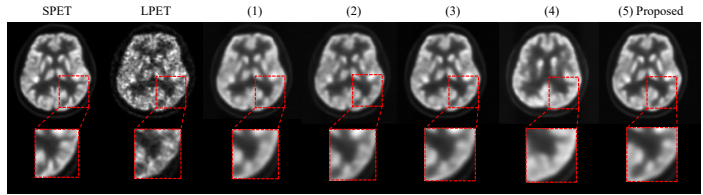

**Fig. 3.** Visual comparison results of the ablation study.

## 4 Conclusion

In this paper, we innovatively propose a multi-modal conditioned adversarial diffusion model to reconstruct high-quality SPET images from the corresponding LPET images with the assistance of the clinical tabular data. We develop a Mc-Encoder that embeds OMTA modules to effectively harness the multi-modal inputs, thereby providing sufficient complementary clues for reconstruction. Meanwhile, the M$^3$TRec is employed to preserve the high-level semantics that are prone to distortion in the diffusion process, further enhancing the reconstruction with accurate semantics. Experimental results on a public dataset have demonstrated the feasibility and superiority of our method. In our future study, we will focus on enhancing the robustness and generalization of our model in complex clinical scenarios with different populations, varying levels of image quality and incomplete data.